# Engineering Diffusivity and Operating Voltage in Lithium Iron Phosphate through Transition Metal Doping

Ajit Jena and B. R. K. Nanda

*Condensed Matter Theory and Computational Lab,*

*Department of Physics, Indian Institute of Technology Madras, Chennai, Tamilnadu, India, 600036*

Density functional calculations are carried out to understand and tailor the electrochemical profile – diffusivity, band gap and open circuit voltage – of transition metal doped olivine phosphate: $LiFe_{1-x}M_xPO_4$ (M = V, Cr, Mn, Co and Ni). Diffusion and hence the ionic conductivity is studied by calculating the activation barrier, $V_{act}$, experienced by the diffusing $Li^+$ ion. We show that the effect of dopants on diffusion is both site dependent and short ranged and thereby it paves ways for microscopic control of ionic conductivity via selective dopants in this olivine phosphates. Dopants with lower valence electrons (LVE) compared to Fe repel the $Li^+$ ion to facilitate its outward diffusion, whereas higher valence electron (HVE) dopants attracts the $Li^+$ ion to facilitate the inward diffusion. From the electronic structure calculation we establish that irrespective of the dopant M, except Mn, the band gap is reduced since the M-$d$ states always lie within the pure band gap. Atomically localized $d$ states of HVE dopants lie above the Fermi energy and that of LVE lie below it. Half-filled Mn-$d$ states undergo large spin-exchange split to bury the dopant states in valence and conduction bands of the pristine system and in turn the band gap remains unchanged in $LiFe_{1-x}Mn_xPO_4$. Baring Mn, the open circuit voltage increases with HVE dopants and decreases with LVE dopants.

I. INTRODUCTION

LiFePO$_4$ (LFP) has been the most widely investigated lithium based cathode material in the last two decades after the reversible Li intercalation in this compound was first demonstrated in 1997 [1]. This Olivine phosphate offers a moderate open circuit voltage (OCV) of ~ 3.5 V [1-5] compatible to the presently available electrolytes. In addition, among all the experimentally synthesized phosphates, LiFePO$_4$ has the best achievable gravimetric capacity of ~170 mAh/g[1-4]. The experimental studies suggest that the activation barrier for the $Li^+$ ion in this compound lies in range 0.1 - 0.6eV [6-10] which is reasonable enough for easy Li diffusion. On the other hand the experimentally measured band gap in this compound is as large as 3.8 eV [11] which is believed to be one of the reasons for hindering the electronic conductivity [12, 13]. The other reason is the presence of atomically localized states, both in the conduction and valence band spectrum, which weakens the electron mobility [14, 15, 16].

For an efficient cathode material the OCV should be optimal to store maximum energy in the cell and at the same time decomposition of the electrolyte is prevented. The activation barrier ($V_{act}$) experienced by the Li$^+$ ion should be weak to enhance the diffusion. However, the weak $V_{act}$ should not lead to instability of the cathode material. Several theoretical [4, 5, 8] and experimental efforts [17-22] have been made to improve the electrochemical profile of LiFePO$_4$. By carrying out a high-throughput *ab initio* calculations, Hautier *et al*. [5] have qualitatively suggested that the electrochemical profile of this compound can be improved via transition-metal doping at the Fe site. In a recent experimental study, it has been shown that 25% 3*d* transition metal doping can enhance the electronic conductivity in LiFePO$_4$ [19]. Even though the results are very few and less conclusive, they open up possibilities to design new materials with optimum electrochemical efficiency through transition metal doping. In this regard atomistic simulations using density-functional theory are the most appropriate tools as not only they are realistic and close to experimental observations, but also they provide a microscopic picture on the cause and the effects.

The electronic conductivity in LiFePO$_4$ is explained by many through the polaron conduction mechanism [12, 23-26]. Here as Li vacates one site, a hole is created in the neighboring Fe as the latter's charge state changes from +2 to +3. There with Li hopping from site to site for the ionic motion, a hole (polaron) moves concurrently to generate the weak electron conductivity [12, 26]. However, such a mechanism is not extended to every member of LiMPO$_4$ family. For example LiNiPO$_4$ does not show polaronic conduction [26]. Recent literatures have correlated the band gap with electronic conductivity in doped compounds. With diluted V doping, the electronic conductivity is enhanced from $10^{-6}$ to $10^{-4}$ [27] and in a subsequent work it is shown that the band gap is decreased by 1.7 eV with 25% of V doping [13]. Similarly increase in electronic conductivity through Mn doping is accompanied by a decrease in band gap [28]. Hence, it deems important to reduce the band gap to better the electrochemical profile of olivine phosphates.

In this paper, we have performed the density functional calculations to study the electronic structure of LiFe$_{1-x}$M$_x$PO$_4$ (M = V, Cr, Mn, Co and Ni; x = 0.125, 0.25, 0.5 and 1.0). To profile the electrochemical behavior, we have calculated (i) the band gap within the frame work of GGA+U exchange-correlation functional, (ii) OCV by comparing the total energies of the lithiated and delithiated sub-phases and (iii) $V_{act}$ using the computationally expensive and more accurate climbing image nudged elastic band (CI-NEB) method [29, 30]. We identify the following three clear trends in our study: (a) Except Mn, doping by other transition metal at the Fe site reduces the band gap; (b) OCV increases with HVE dopants (Co and Ni) and decreases with the LVE dopants (V and Cr) and (c) the effect of the dopant on the activation barrier is site selective and short ranged. If the dopant is LVE it pushes the Li$^+$ ion from its immediate neighborhood to diffuse outward and reverse is the case for HVE dopants.

## II. COMPUTATIONAL METHODOLOGY

Quantum espresso (QE) [31] simulation package is used to perform the spin polarized DFT and CI-NEB calculations. DFT results are obtained using the Vanderbilt ultra-soft pseudo-potentials and plane wave basis sets. The kinetic energy cutoff to fix the number of plane waves is taken as 30 Ry. A 6X10X12 k-mesh of the BZ for the regular unit cell is found to be sufficient to calculate the total energy with reasonable accuracy. We have used 2X2X1 and 1X2X1 supercells for 12.5% and 25% doping of the transition metal elements at the Fe site respectively and accordingly appropriate k-mesh are used for self-consistent calculations. The constructed doped structures are further relaxed to achieve the ground state. Since experimental structural parameters for LiVPO$_4$ and LiCrPO$_4$ are not available, we took the structure of LiFePO$_4$ [32] as the initial one and optimized it subsequently. To account for the strong correlation effect, parameterized Hubbard U [33] (= 3 eV) is included in our calculations for all the transition metal dopants. The values of OCV and band gap are sensitive to the value of U [34]. In our earlier work [14] we have shown that the insulating behavior of this compound arises due to atomically localized transition metal $d$ states and U simply amplifies the gap. We find that for U = 3 eV, the OCV of LiFePO$_4$ is 3.21 V, which is close to experimental value of 3.5 V. Based on the Eq. 3, which will be discussed later, expression of OCV includes total energy for lithiated and delithiated phases. The trend of OCV, across the transition metals as well as doping concentrations, will remain same irrespective of the value of U.

To study the Li$^+$ ion diffusion in doped LFP, we have employed the CI-NEB [29, 30] method based on the transition state theory [35]. The minimum energy path (MEP) or the diffusion path for the conducting Li$^+$ ion was obtained using NEB algorithm which is illustrated in Fig. 1. First an initial (guess) path connecting a set of equi-spaced images is made. Here the final image point represents a Li vacancy site to which the Li$^+$ ion from the initial image point will hop via the intermediate image points. The force, experienced by the Li$^+$ ion at each image point, $i$, $F_i = F_i^{s\|} + F_i^{\nabla_\perp}$. While the spring force $F_i^{s\|}$ keeps the images equi-spaced, the true force $F_i^{\nabla_\perp}$ displaces the intermediate images. The MEP is realized when $F_i^{\nabla_\perp}$ becomes zero. The MEPs obtained for LiFe$_{1-x}$M$_x$PO$_4$ are shown in Fig. 1(c). It repeats the well-known non-linear one dimensional path of LFP [36] irrespective of M and x. To provide a quantitative measure to the diffusion, we have calculated the V$_{act}$ using CI-NEB method. The activation barrier is defined as the potential energy difference between the saddle point and the initial image (forward V$_{act}$) or final image (backward V$_{act}$). The CI-NEB method, while retains the MEP obtained from NEB, it improves the accuracy of the potential energy at the saddle point [30]. The diffusivity D is related to $V_{act}$ by $D = D_0 \exp\left(-\frac{V_{act}}{kT}\right)$, where $D_0$ is the diffusion constant which depends on the hopping length and $k$ is the Boltzmann constant [35, 37].

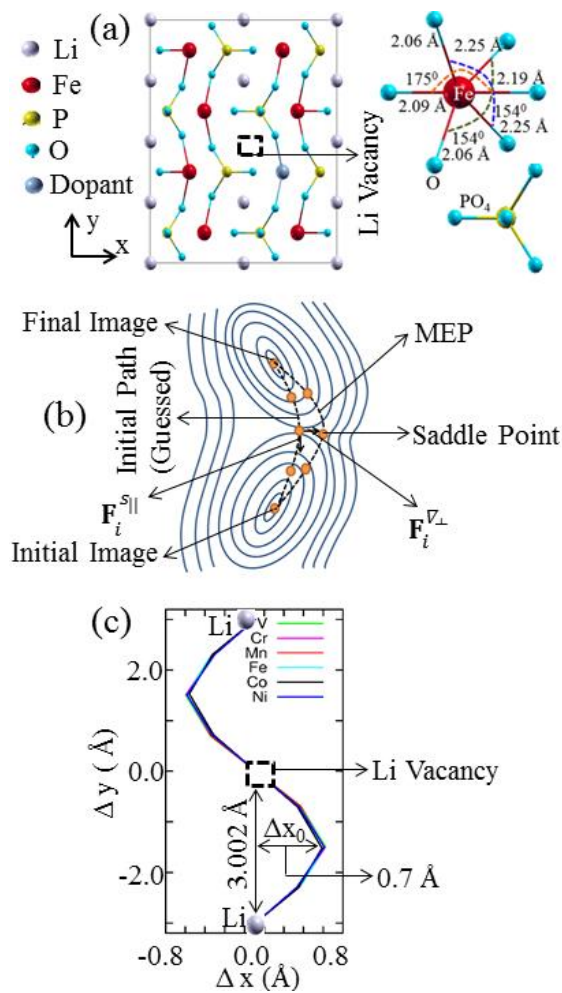

FIG. 1. (a) LiFePO$_4$ unit cell viewed from 001 plane. The crystal structure has perfect PO$_4$ tetrahedras and asymmetric FeO$_6$ complexes. The open square indicates a vacancy. When a vacancy is created, neighboring Li$^+$ ion hops to the vacant cite and initiate the diffusion process. (b) Schematic illustration of NEB method and equipotential contours appropriate for olivine phosphates [36]. (c) NEB Calculated diffusion path for the Li$^+$ ion in LiFe$_{1-x}$M$_x$PO$_4$. Here ($\Delta x$, $\Delta y$) represents the coordinate of the hopping Li ion during motion with respect to the vacancy site.

## III. BANDSTRCUTURE AND OPEN CIRCUIT VOLTAGE

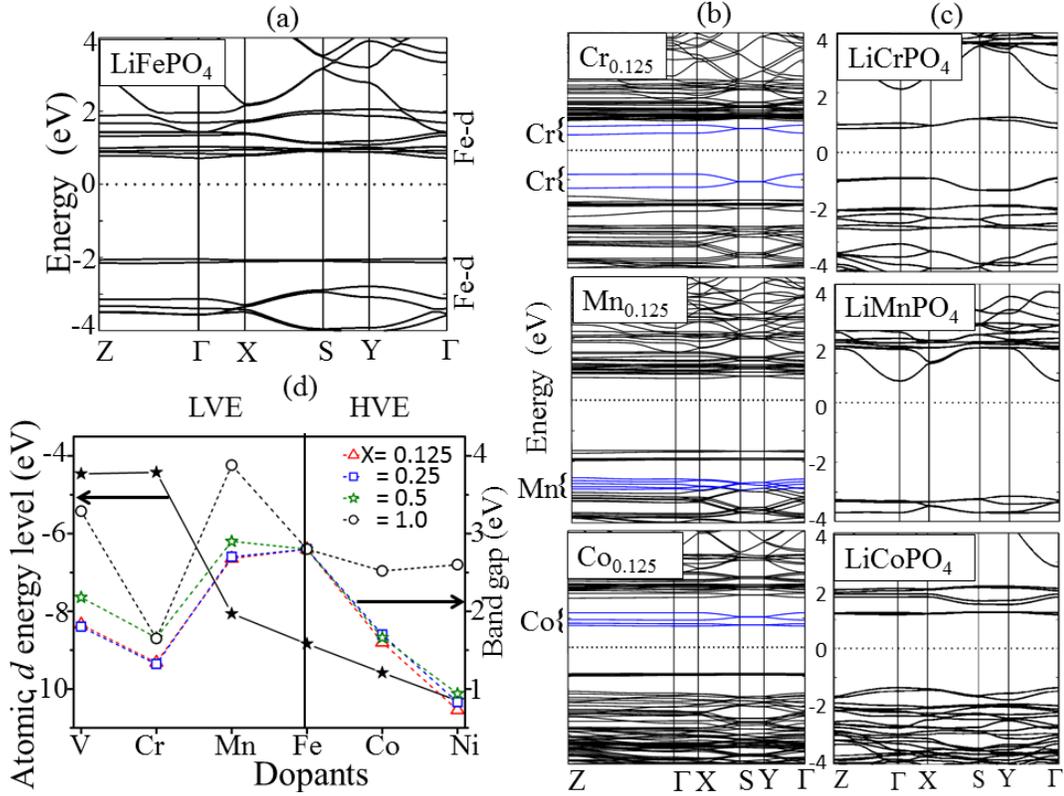

FIG. 2. (a), (b) and (c) respectively show the ground state band structure of LiFePO$_4$, LiFe$_{1-x}$M$_x$PO$_4$ (M = Cr, Mn, Co; x = 0.125) and LiMPO$_4$. (d) The solid line with filled stars shows the dopant atomic $d$-energy levels $E_d = \left[E\left(d_{\frac{3}{2}}\right) + E\left(d_{\frac{5}{2}}\right)\right]/2$. We note that the difference between $E\left(d_{\frac{3}{2}}\right)$ and $E\left(d_{\frac{5}{2}}\right)$ is negligible and of the order of 0.2 eV. Since we are interested in examining the relative energy position of the atomic $d$ states the average energy is considered. The dashed lines in (d) show the band gaps of LiFe$_{1-x}$M$_x$PO$_4$ for different concentrations (x). The Fermi energy is set to zero.

The DFT+U band structure of LFP is shown in Fig. 2a. A detailed analysis of it is presented in one of our recent work [14]. The electronic and magnetic structure of LFP are governed by FeO$_6$ complex and PO$_4$ tetrahedra. The symmetry in the metal-oxygen complex in LFP is highly distorted and creates a completely anisotropic crystal field for the Fe-$d$ states leading to loss of threefold $t_{2g}$ and twofold $e_g$ degeneracy. As a consequence we have atomically localized $d$-states which make the system insulating. The narrow gap produced through this localization gets amplified due to strong correlation effect as in the case of Mott insulators. Hence, LFP is considered as weakly coupled Mott insulator. To study the change in the electronic structure with doping we have plotted the band structure of LiFe$_{0.875}$M$_{0.125}$PO$_4$ and LiMPO$_4$ in Fig. 2b and c respectively for selected dopants (M = Cr, Mn and Co). The significant changes occurred is that in the case of Cr and Co doping, new states (colored lines) are found lying within the band gap, either below or above $E_F$, of the pure compound. This is true for all other dopants (not shown here), except Mn. From calculations of partial densities of states, these new states are identified as dopant M-$d$ states. Therefore, the band gap is always reduced with doping irrespective of M and barring Mn. The Mn-$d$ states, as shown in the middle panel of Fig. 2b, are buried inside the Fe bands.

The reason for band gap reduction with doping can be understood from the comparison between the band structure of the doped compounds and the corresponding pristine compounds LiMPO$_4$ which are plotted in Fig. 2c. We find that the energy level of top valence bands and bottom conduction bands of LiMPO$_4$ nearly matches with that of the dopant $d$-states in LiFe$_{1-x}$M$_x$PO$_4$. This is due to the fact that even though the Olivine phosphates are crystalline solids, the Bloch $d$ electrons behave like atomically localized electrons [14]. To substantiate it further we have plotted the atomic M-$d$ energy levels in Fig. 2d. As expected, V and Cr-$d$ energy levels are higher to that of Fe-$d$. Following the same trend, the dopant V and Cr-$d$ states appear above the valence-$d$ states as can be seen from the upper panel of Fig. 2b. As a consequence, LVE dopants reduce the band gap. Similarly Ni and Co-$d$ energy levels are lower to that of Fe-$d$. Therefore, while the Ni and Co-$d$ valence states are deep inside the valence band spectrum, their conduction states are below the Fe-$d$ conduction states and hence the band gap is reduced here as well.

In addition to the atomic $d$-energy levels, the strength of the spin-exchange splitting in this family of antiferromagnetic and insulating Olivine phosphates has an important role in positioning the M-$d$ states. Since Mn$^{2+}$ ($3d^5$) is half-filled, the exchange splitting is very large and hence the system shows a wide band gap compared to the other members. The band structure of LiMnPO$_4$, calculated within GGA predicts a band gap of 2.2 eV and with U (= 3eV) it becomes 3.89 eV (see Fig. 2d). Therefore, even though atomic Mn-$d$ levels are comparatively higher than the atomic Fe-$d$ levels, in the doped system LiFe$_{1-x}$Mn$_x$PO$_4$, Mn-$d$ valence (conduction) states lie below (above) the Fe-$d$ valence (conduction) states. As a consequence, band gap remains almost unchanged with Mn doping. The minor deviation is attributed to the shift in the Fe-$d$ states. On the other hand, Cr$^{2+}$ ($3d^4$) has relatively weak spin-exchange split compared to Fe and hence the Cr-$d$ valence and conduction states lie above and below the Fe-$d$ valence and conduction states to reduce the band gap substantially. The band gap of LiFe$_{1-x}$M$_x$PO$_4$ (M = V, Cr, Mn, Co, Ni; x = 0.125, 0.25, 0.5 and 1) plotted in Fig. 2d augur well with this understanding. The reduction in the band gap is not restricted to doped LiFePO$_4$. Our studies reveal that the band gap of any transition metal doped LiMPO$_4$ is always smaller than that of the respective pure compound. Even though the results are not presented here in details, this still can be observed in the case 50% doping. The band gap of LiFe$_{0.5}$M$_{0.5}$PO$_4$ is always less than that of LiMPO$_4$.

One of the advantages of LiFePO$_4$ as the cathode material lies in its structure. As shown in Fig. 1(a), the Li and Fe atoms can be assumed to be distributed in a matrix of PO$_4$ tetrahedras. Since the stability of the system is mostly related to these tetrahedras [14], Li$^+$ ion can be easily inserted in Fe$_{1-x}$M$_x$PO$_4$ during lithiation and extracted out of LiFe$_{1-x}$M$_x$PO$_4$ during delithiation [1, 38-40]. The lithiation and delithiation process can be expressed respectively as[1]:

$$Fe_{1-x}M_xPO_4 + yLi^+ + ye^- \rightarrow yLiFe_{1-x}M_xPO_4 + (1-y)Fe_{1-x}M_xPO_4. \qquad (1)$$

$$LiFe_{1-x}M_xPO_4 \rightarrow yLi^+ + ye^- + (1-y)LiFe_{1-x}M_xPO_4 + yFe_{1-x}M_xPO_4. \qquad (2)$$

Therefore, the intercalation or the open circuit voltage [41, 42] which is also the maximum possible operating voltage, is

$$OCV = E(Fe_{1-x}M_xPO_4) + E(Li) - E(LiFe_{1-x}M_xPO_4). \qquad (3)$$

Here, E denotes the total energy of the corresponding system which is obtained from the density-functional calculations. We have considered the experimental BCC crystal structure of Li [43] to calculate E(Li) and the optimized structure of (Li)Fe$_{1-x}$M$_x$PO$_4$ to calculate E((Li)Fe$_{1-x}$M$_x$PO$_4$). The OCV of LiFe$_{1-x}$M$_x$PO$_4$ (x = 0.125, 0.25, 0.5, 1.0) are shown in Fig. 3a.

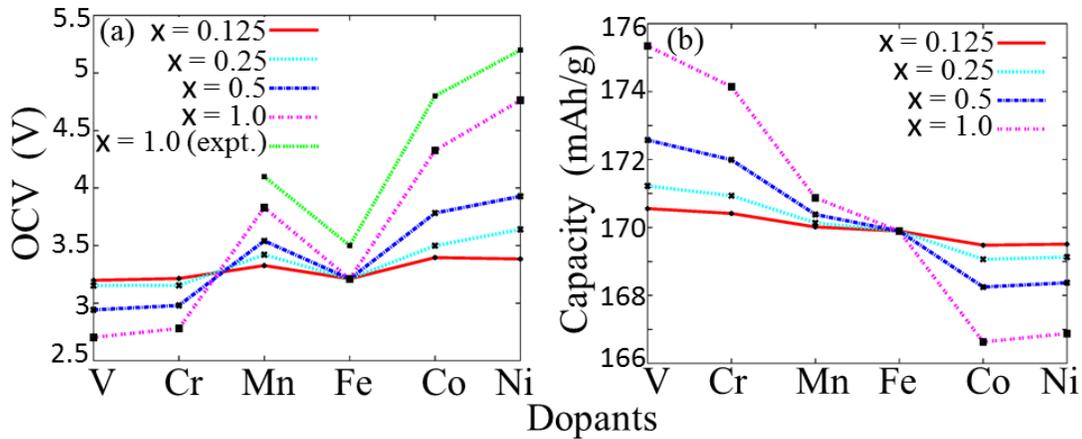

FIG. 3. (a) OCV and (b) gravimetric capacity of LiFe$_{1-x}$M$_x$PO$_4$. While OCV is calculated using Eq. 3, the gravimetric capacity is obtained using Faraday's relation [44]. The green dashed line in (a) shows the OCV obtained from experimental studies [1, 38-40]. For a given dopant concentration there always exist more than one possible configuration. However, we have found that all these configurations give almost same OCV and therefore, we have shown the results for a single configuration for each dopant to avoid redundancy.

From Fig. 3a we gather that DFT underestimates the OCV compares to the experimental results. However, the trend among the pure compounds remains same. The minor underestimation of OCV by DFT is a known fact as several other studies listed in table-I reveal the same. The table also shows that the calculations carried out with GGA+U approximation provides a better estimation of OCV than that of GGA. Within GGA, the strong correlation effect on the transition metal-*d* states is not accounted appropriately. As a consequence the destabilization energy of both non-lithiated and lithiated compounds is underestimated. However, the underestimation is far more for the non-lithiated compound compared to that of the lithiated compound [41, 42]. Therefore, based on Eq. 3, GGA provides lower value of OCV. Calculations carried out using LDA show much smaller OCV [41, 42, 45].

As far as the OCV of doped LFP is concerned, from Fig. 3, we find that, except Mn, the LVE dopant decreases the OCV and HVE dopant increases it. This trend resembles the OCV of the pristine compounds. For completeness we have estimated the gravimetric capacity using the Faraday's relation: $C = \frac{nF}{3600 \times M_w} \, mAh/g$, where $n$ is charge of the Li ion, $F$ is the Faraday's constant and $M_w$ is the molecular weight. As expected, the capacity increases with LVE dopants and decreases with HVE dopants (see Fig. 3b). For Ni, the capacity is little larger than that of Co as the former has lighter atomic mass than the latter. Fig. 3(a) and (b) together imply that OCV and gravimetric capacity are inversely proportional in LiMPO$_4$, which is an unexplored relation to our knowledge. This implies that in LiMPO$_4$, instability increases with lighter transition metal elements.

TABLE I. Experimental and theoretical values of OCV (in volts) of LiMPO$_4$. Besides our results, data from literatures are also listed to make a comparison.

| M | OCV (GGA) | | OCV (GGA +U) | | OCV |
| --- | --- | --- | --- | --- | --- |
| | Present work | Literature | Present work | Literature | (experiment) |
| Mn | 2.94 | 2.98[45] | 3.83 | 4.04[45], U = 3.92 eV | 4.1 |
| Fe | 2.52 | 2.99[45] | 3.21 | 3.47[45], U = 3.71 eV | 3.5 |
| Co | 3.63 | 3.64[42], 3.70[45] | 4.33 | 4.73[45], U = 5.05 eV | 4.8 |
| Ni | 4.29 | 4.20[45] | 4.76 | 5.07[45], U = 5.26 eV | 5.2 |

## IV. Li$^+$ ION DIFFUSION

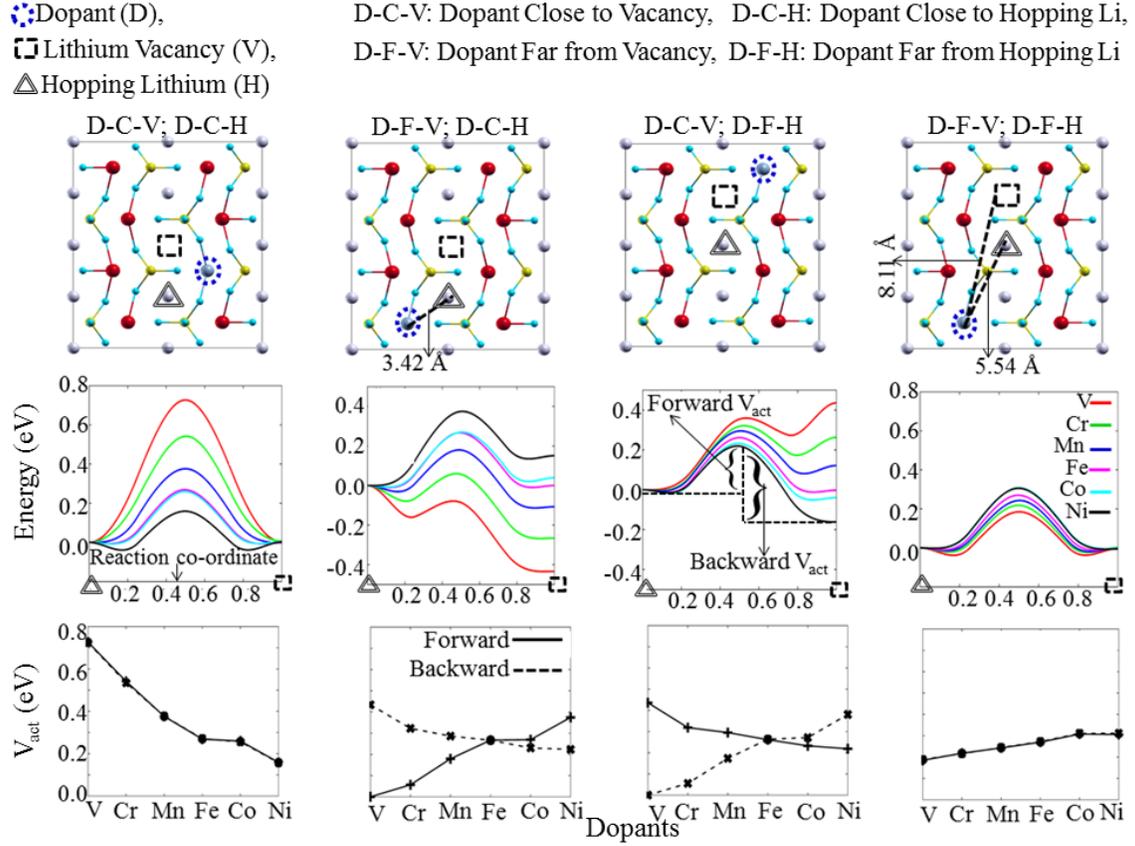

FIG. 4. Upper panel: Several configurations representing different position of the dopant (D) with respect to the hopping Lithium site (H) and vacancy (V). The sites H and V respectively represent initial and final image points in the CI-NEB method discussed earlier. The configuration [D-C-V; D-C-H] or [D-F-V; D-F-H] stands for dopant close to or far from both V and H. Middle panel: The potential barrier experienced by the hopping Li$^+$ ion across the diffusion path as shown in Fig. 1c for the corresponding configuration of the upper panel. Lower panel: Estimated forward and backward $V_{act}$. The results in the present work are obtained using the non-magnetic configuration. Since change in the total energy with change in magnetic ordering is of the order of meV, it is expected that the physical parameters like $V_{act}$ and OCV are nearly independent of magnetic ordering in LiFePO$_4$.

The utility of the LFP largely depends on Li$^+$ diffusion since the latter determines the ionic conductivity in this system. To investigate the effect of dopants on Li$^+$ diffusion, we have calculated the $V_{act}$ using CI-NEB method discussed earlier in computational methodology, for different dopants and dopant neighborhoods. Here we present four cases as shown in Fig-4: (a) dopant is close to both the vacancy and the hopping site [D-C-V; D-C-H], (b) dopant is far from vacancy but close to hopping site, [D-F-V; D-C-H], (c) dopant is close to vacancy and away from hopping site [D-C-V; D-F-H], and (d) dopant is away both from vacancy and hopping site [D-F-V; D-F-H]. The corresponding potential barrier across the diffusion path and the resulted $V_{act}$ are shown on the respective columns of Fig. 4.

First to validate our results, in table-II, we have made a comparison of $V_{act}$ of LiMPO$_4$ with the available experimental and theoretical values. Experimentally $V_{act}$ is measured by several techniques such as *ac* and *dc* impedance measurements, Mossbauer spectroscopic and *μ-S-R* measurements. As expected, most of the studies are done on LiFePO$_4$. While the *μ-S-R* measurements report a very low value of $V_{act}$ (~ 0.1 eV) [6, 46], the other techniques report a wide range of values between 0.15 and 0.66 [7, 9, 10, 47]. One of the earlier NEB study on LiMPO$_4$ family of compounds estimates the $V_{act}$ in the range 0.13 to 0.36 [48]. It is maximum for M = Co and minimum for M = Ni. Atomistic simulations, using a model parametric Buckingham potential, carried out by Fisher *et al*.[49], provide higher value of $V_{act}$ ( 0.44 – 0.62). The latter also shows a definitive trend, i.e., $V_{act}$ decreases as we move from LVE to HVE transition metal element. Our calculations for the configuration [D-C-V; D-C-H], provide a similar trend. The $V_{act}$ is maximum for M = V (0.72 eV) and minimum for M = Ni (0.16 eV). While our results are for the doped systems instead of LiMPO$_4$, they certainly reveal that when dopant is close to both vacancy and hopping lithium site, it replicates the un-doped system. This significantly infers that the effect of dopant on diffusion is very short ranged.

To further verify the short range behavior, we have examined a configuration [D-F-V; D-F-H] where the dopant is away both from both vacancy and hopping site (see Fig. 4). The resulted $V_{act}$ changes very little with dopant and lies in the range 0.2 to 0.3 eV unlike the range 0.16 to 0.72 seen in the case of [D-C-V; D-C-H]. Therefore, the effect of dopant on the diffusion rapidly diminishes with increase in the separation between dopant and diffusing Li. Earlier it has been discussed that the bands in LiFe$_{1-x}$M$_x$PO$_4$ are nearly non-dispersive near $E_F$. These localized bands are the outcome anisotropic crystal field in the system. The asymmetric crystal field splits the metal-*d* states into five non-degenerate states. Therefore, while the olivine phosphates are crystalline, their electronic properties resemble to that of atoms. The complex PO$_4^{3-}$ is one of the most stable polyanions involving phosphorous and oxygen and the stability is attributed to the ionic bonding between them. The other interactions, Fe$^{2+/3+}$- Li$^+$, O$^{2-}$- Li$^+$ and L$^+$-P$^{5+}$, are comparatively negligible. Hence, the effect of dopant on diffusion is very short ranged.

To realize whether both short range effect and electronic behavior of dopant can be utilized to manipulate the diffusion microscopically, we have estimated $V_{act}$ as a function of dopant for two new configurations [D-F-V; D-C-H] and [D-C-V; D-F-H] shown in Fig. 4. From the results, shown in the middle panel, we infer the following. (a) Unlike the earlier two configurations, here the potential barrier across the diffusion path is asymmetric with respect to forward and backward motion (see third row of Fig. 4). In the forward motion the Li$^+$ ion moves from H to V. In the other case, the motion is reversed since a new vacancy is created at the original hopping site. (b) In the case of [D-F-V; D-C-H], the potential is down-hill for the forward motion and up-hill for the backward motion for the LVE dopants. Accordingly the forward $V_{act}$ is less and the backward $V_{act}$ is more. For HVE dopants the forward $V_{act}$ is more and backward $V_{act}$ is less. The situation reverses for [D-C-V; D-F-H]. As a whole, our results conclude that when the diffusing Li is in close proximity to the LVE (HVE) dopant, a repulsive (attractive) force acts on it to facilitate outward (inward) diffusion. This opens up opportunities and future research prospect to tune the diffusivity through intentional inhomogeneous doping. The affinity between Li$^+$ ion and transition element is governed by the number of valence electrons of the latter. With more valence electrons, the *d*-states of the transition metal lie lower in energy and offer a higher attractive potential to Li$^+$ ion.

TABLE II. Comparison of $V_{act}$ of LiMPO$_4$ as obtained from different experimental and theoretical techniques. We also have listed the results from our CI-NEB calculations on LiFe$_{0.875}$M$_{0.125}$PO$_4$ for the configuration [D-C-V; D-C-H] (see Fig.4).

| M<br><br>Activation Barrier (eV) | V | Cr | Mn | Fe | Co | Ni |
| --- | --- | --- | --- | --- | --- | --- |
| Experiment | -- | -- | 0.65 - 1.14 * | 0.155 – 0.66*<br><br>0.18 – 0.53**<br><br>0.1 *** | 0.62*<br><br>0.1*** | 0.61*<br><br>0.17*** |
| Present Work(CI-NEB) | 0.72 | 0.54 | 0.38 | 0.27 | 0.26 | 0.16 |
| Literature (NEB) [48] | -- | -- | 0.25 | 0.27 | 0.36 | 0.13 |
| Literature (Buckingham Model Potential) [49] | -- | -- | 0.62 | 0.55 | 0.47 | 0.44 |

*AC impedance measurement [7, 9, 10, 47]; ** DC impedance measurement [8, 10]; ***μ-S-R measurement [6, 46]

## V. SUMMARY AND CONCLUSIONS

In summary, we present a detail electronic and electrochemical profile of transition metal doped LiFePO$_4$ by carrying out density-functional and CI-NEB simulations. Our results provide the following interesting conclusions: (a) Doping leads to reduction in the band gap irrespective of the nature of the dopant, barring Mn, and the reason is attributed to the atomically localized nature of the Bloch *d* electrons in this family; (b) the open circuit voltage increases with increasing valence electrons in the dopant; (c) the effect of dopant on the Li$^+$ ion diffusion is very short ranged; and (d) while LVE dopants repel the Li$^+$ ion, the HVE dopants tend to attract it. Our study, therefore, opens up the options for optimizing the diffusivity in particular and electrochemical behavior in general for LiFePO$_4$ through tailored transition metal doping. We expect that the conclusions made in this work can be extended to other Li based cathode materials where Li has weak interaction with the transition metals.

## ACKNOWLEDGEMENTS

This work was funded by NISSAN RESEARCH PROGRAM through grant no. PHY1314289NRSPBIRA. Institute HPCE facility was used for the computations. The authors would like to thank Sudakar Chandran for stimulating discussions.